# Cultivating Cybersecurity: Designing a Cybersecurity Curriculum for the Food and Agriculture Sector

George Grispos[1], Logan Mears[1], Larry Loucks[2], William Mahoney[1]
[1]University of Nebraska-Omaha, USA
[2]Ponca Economic Development Corporation, USA

ggrispos@unomaha.edu
lmears@unomaha.edu
lloucks@pedco-ne.org
wmahoney@unomaha.edu

**Abstract**[1]**:** As technology increasingly integrates into farm settings, the food and agriculture sector has become vulnerable to cyberattacks. However, previous research has indicated that many farmers and food producers lack the cybersecurity education they require to identify and mitigate the growing number of threats and risks impacting the industry. This paper presents an ongoing research effort describing a cybersecurity initiative to educate various populations in the farming and agriculture community. The initiative proposes the development and delivery of a ten-module cybersecurity course, to create a more secure workforce, focusing on individuals who, in the past, have received minimal exposure to cybersecurity education initiatives.

**Keywords:** Cybersecurity, Farming, Agriculture, Education, Workforce Development

## 1. Motivation

The food and agriculture sector, classified as critical infrastructure in the United States (U.S.), ensures food security and economic stability for large parts of the country (Versteeg, 2023). However, integrating technology into farm settings has left the sector vulnerable to cyberattacks. For example, a ransomware attack in 2021 resulted in a fifth of the beef processing plants in the U.S. to shut down, with one organization also paying an $11 million ransom (Bunge, 2021). Similarly, an attack impacting a grain storage cooperative in Iowa caused operational disruptions and data theft (Plume and Bing, 2021).

In 2010 the U.S. Department of Agriculture classified cybersecurity as a low priority for the food and agriculture industry sector. While this was reversed in 2015 (Internet Security Alliance, n.d.), the damage was already, potentially done. A 2018 survey reported that "a concerning number of farmers either do not have computer security applications installed or do not know whether computer security applications are installed on their computers" (Spaulding and Wolf, 2018). Likewise, a report from the Department of Homeland Security (2018) added that many farmers do not take cybersecurity risks seriously enough. Recently, the Cybersecurity and Infrastructure Security Agency (2022) suggested that ransomware attacks against the industry could have been prevented with the implementation of basic cybersecurity practices. Hence, as malicious actors look to exploit technologies in farm settings, coupled with limited cybersecurity awareness in the food and agriculture sector, there is a growing need to train farmers on basic cybersecurity hygiene measures.

This paper presents an ongoing effort towards the development and delivery of a cybersecurity curriculum designed to help increase cybersecurity awareness among farmers and food producers, who in the past (Geil et al., 2018), have received minimal exposure to cybersecurity education initiatives. The goal of the curriculum is to help create a more secure workforce in the food and agricultural sector. The remainder of this work-in-progress paper is structured as follows. Section two positions the current work in terms of related work, while Section three presents the proposed curriculum. Section four concludes the paper and discusses the delivery of the curriculum to the food and agricultural sector.

---

[1] Please cite this pre-print as: G. Grispos, L. Mears, L. Loucks, W. Mahoney (2025). Cultivating Cybersecurity: Designing a Cybersecurity Curriculum for the Food and Agriculture Sector, 20th International Conference on Cyber Warfare and Security (ICCWS 2025), Williamsburg, Virginia, USA

## 2. Related Work

While various technologies are being integrated into farm settings, these technologies introduce several cybersecurity concerns for the food and agricultural community. Demestichas, et al (2020) surveyed technical cybersecurity threats in smart farming and discussed their impact on the agricultural sector. Freyhof, et al. (2025) investigate the risks associated with cyberattacks impacting CAN bus-driven farming equipment, reporting that a farmer would sustain financial losses, if such attacks are successful. Gupta et al. (2020), add that Internet of Things devices integrated into farms are not being designed with security in mind and could be compromised for "agri-terrorism" activities. Other researchers have focused their efforts on technical solutions to these problems including the proposal of a cybersecurity framework for farming (Chi et al., 2017), the inclusion of cybersecurity policies (Barreto and Amaral, 2018), and the development of cybersecurity testbeds (Freyhof et al., 2022).

Regardless of the approach, previous research has highlighted that many farmers and food producers are not aware of the risks associated with implementing technology on their farms (Nikander et al., 2020, Dehghantanha et al., 2021). Further complicating matters, Spaulding and Wolf (2018) investigated cybersecurity awareness among farmers and reported that approximately 10% of farmers had attended previous cybersecurity training. As a result, many farmers lack the education they require to identify and mitigate cyber threats on their farms. This is a view shared by Geil et al. (2018), who added that "there is a need for computer security education within the industry" and that an opportunity exists for organizations to offer a cybersecurity education program.

In the past, various cybersecurity educational initiatives have been developed for different industries and settings, including critical infrastructure. Waddell, et al (2024) outline a cybersecurity program developed for healthcare, which is based on strategies previously adopted in commercial aviation, including dynamic education delivery options, and focused simulations. Chowdhury and Gkioulos (2021) conducted a systematic literature review that investigated cybersecurity training specially designed for critical infrastructure settings. One of the main findings from this analysis was that many initiatives tend to focus on training scenarios and team-based exercises, rather than traditional informational methods. Kessler and Ramsay (2013) argue that while many colleges and universities in the U.S. have cybersecurity programs, many of these programs are not suitable for the homeland security domain. Hence, Kessler and Ramsay (2013) propose a number of ideas that could be integrated into a cybersecurity program that can help homeland security students. While previous research has looked to develop various cybersecurity educational initiatives for critical infrastructure, minimal research has investigated the development and delivery of a cybersecurity program for the food and agricultural sector.

## 3. Proposed Curriculum

The curriculum consists of ten modules (as shown in Table 1), that will be delivered using an online learning platform. Each module consists of open-source readings and classroom instruction-style videos that are thirty minutes long and include case studies and news articles where appropriate.

| Module | Description |
| --- | --- |
| *Introduction to Cybersecurity* | Defines common cybersecurity terms and presents cybersecurity nomenclature. Topics include the CIA triad and cybersecurity risk. |
| *Computer Networks and the Internet* | Discusses the structure of the Internet and introduces network concepts and devices. Topics include LANs, WANS, wireless networks, and networking devices. |
| *Social Engineering and Other Attacks* | Defines and discusses social engineering attacks and examines how Internet fraud and identity theft works. Topics include social engineering attacks and countermeasures of these type of attacks. |
| *Malware and Ransomware* | Defines and discusses malware and ransomware attacks and how to defend against such attacks. Topics include the impact of malware, different types of malware, as well as countermeasures for preventing and recovering from malware attacks. |
| *Computer Security Software* | Discusses virus scanners, firewalls, and intrusion detection systems. Topics include why computer security software is important and how to obtain and setup this software in farm settings. |
| *Mobile Device and Wireless Security* | Discusses mobile device security (including applications), threats using open networks, and hardening access points. Topics include securing a mobile device, threats related to mobile devices, and cybersecurity threats associated with wireless networks. |
| *Cyber Warfare* | Discusses types of cyber warfare, examines motivations for cyber warfare, and presents relevant case studies. This includes case studies of previous cyberattacks from a cyber warfare perspective. |
| *Responding to Incidents and Problems* | Introduces how to respond to security incidents, including basic first responder actions. Topics include steps for responding to security incidents as well as first responder actions. |
| *Farm and Food Cybersecurity Act* | Discusses the proposed Act, and its implications on the sector. |
| *Countermeasures* | Presents best practices and countermeasures that can be implemented to help enhance the cybersecurity posture of a farm. |

**Table 1. Module Names and Descriptions**

To ensure that the curriculum will "help safeguard America's national security and economic prosperity" (National Institute of Standards and Technology, 2016), the modules have been developed and mapped to the NICE Cybersecurity Workforce Framework. While the NICE Framework does not include a work role associated with the agriculture industry, the modules cover the following knowledge areas and skills:

- **Knowledge Areas**: K0001, K0002, K0003, K0004, K0005, K0006, K0011, K0021, K0026, K0033, K0040, K0041, K0042, K0044, K0049, K0057, K0059, K0060, K0104, K0106, K0111, K0113, K0114, K0138, K0147, K0150, K0151, K0160, K0165, K0167, K0174, K0181, K0205, K0210, K0211, K0219, K0270, K0274, K0283, K0324, K0327, K0332, K0340, K0344, K0345, K0362, K0392, K0397, K0443, K0444, K0471, K0480, K0487, K0515, K0523, K0536, K0555, K0561, K0565
- **Skills**: S0003, S0006, S0040, S0059, S0079, S0081, S0084, S0086, S0154, S0168

The curriculum's learning outcomes are based on two categories, from Bloom's Taxonomy (Armstrong, 2010): *remember, understand* and, *apply*. For example, after completing the 'Introduction to Cybersecurity' module, individuals should be able to *remember* and *understand* the principles of confidentiality, integrity, and availability, while after completing the "Malware and Ransomware" module, individuals should be able to *apply* and implement basic cyber hygiene practices to minimize the impact of these threats.

## 4. Curriculum Delivery and Future Work

The next step is to target the curriculum to different populations in the food and agriculture sector. The first population is students who are currently enrolled in agriculture degree programs, i.e. tomorrow's farmers and food producers. According to a U.S. Government report (McFadden et al., 2023), farmers with less than ten years of experience are more likely to adopt technology on their farms. Hence, targeting this population for exposure to the curriculum supports these individuals towards creating a more secure environment, if they choose to implement technology in their future farm settings. Agriculture degree programs in the University of Nebraska System will be leveraged to identify students who are interested in learning more about cybersecurity, who will then be subjected to the curriculum.

The second population is farmers and food producers in general. The University of Nebraska has a track record concerning educational initiatives for farmers and food producers "already in the field". These cover a variety of topics from improving general farming practices and techniques, to the use of technologies (e.g. drones) in farm settings (UNL Nebraska Extension, n.d.). The cybersecurity curriculum will form part of these initiatives, where farmers in Nebraska will be invited to attend a workshop that will be used to deliver the module content. While the curriculum has been designed for online learning, the same content can also be delivered in a face-to-face setting.

While farmers throughout the U.S. could be impacted by a cyberattack, there is the potential that attacks against American Indian farmers could have wider consequences for not the farmer, but the broader tribe. According to the USDA (2017), there are 5,037 farms in the U.S. owned by American Indians, with 200 of these in Nebraska, many of which belong to the Ponca Tribe. Ponca Farmers specifically produce and sell crops to a grocery store located in a "food desert" (Nebraska Legislature, n.d. , Dockendorf, 2020). As a result, potential cyberattacks impacting these farmers could have dire food security consequences for not only the Tribe but an entire area. American Indian farmers will be targeted for exposure to the cybersecurity curriculum as "critical" farmers and food producers.

Regardless of the population, to measure the effectiveness of the curriculum, a *pretest-treatment-posttest* experiment will be undertaken (Campbell and Stanley, 1963). Before being subjected to the *treatment* (the curriculum), the *subjects* (farmers) will be given a *pretest survey* (measurement), to establish the subject's cybersecurity knowledge before being exposed to the treatment. Similarly, after exposure to the treatment, the subjects will be given a *posttest survey*, to determine if the subject's cybersecurity knowledge has increased or decreased because of exposure to the treatment.

## Acknowledgements

This research was support by the U.S. National Science Foundation (NSF) through the Award #2335812: "Education DCL: EAGER: Enhancing Cybersecurity Awareness of American Indian Farmers and Food Producers: The Ponca Tribe of Nebraska as a Case Study". The statements, opinions, and content included in this publication do not necessarily reflect the position or the policy of the NSF, and no official endorsement should be inferred.